\documentclass
[a4paper,final,showpacs,prl,superscriptaddress,amsfonts,footinbib,amssymb,twocolumn]{revtex4}%
\usepackage{amsfonts}
\usepackage{amsmath}
\usepackage{amssymb}
\usepackage{graphicx}%
\providecommand{\U}[1]{\protect\rule{.1in}{.1in}}
\providecommand{\U}[1]{\protect\rule{.1in}{.1in}}

\begin{document}
\title{Quantum parallelism as a tool for ensemble spin dynamics calculations}
\author{Gonzalo A. \'{A}lvarez}
\email{galvarez@famaf.unc.edu.ar}
\affiliation{Facultad de Matem\'{a}tica, Astronom\'{\i}a y F\'{\i}sica, Universidad
Nacional de C\'{o}rdoba, 5000 C\'{o}rdoba, Argentina.}
\author{Ernesto P. Danieli}
\affiliation{Macromolecular Chemistry, RWTH Aachen, Sammelbau Chemie, Worringer Weg 1,
D-52056 Aachen, Germany.}
\author{Patricia R. Levstein}
\affiliation{Facultad de Matem\'{a}tica, Astronom\'{\i}a y F\'{\i}sica, Universidad
Nacional de C\'{o}rdoba, 5000 C\'{o}rdoba, Argentina.}
\author{Horacio M. Pastawski}
\email{horacio@famaf.unc.edu.ar}
\affiliation{Facultad de Matem\'{a}tica, Astronom\'{\i}a y F\'{\i}sica, Universidad
Nacional de C\'{o}rdoba, 5000 C\'{o}rdoba, Argentina.}

\begin{abstract}
Efficient simulations of quantum evolutions of spin-$1/2$ systems are relevant
for ensemble quantum computation as well as in typical NMR experiments. We
propose an efficient method to calculate the dynamics of an observable
provided that the initial excitation is \textquotedblleft
local\textquotedblright. It resorts a single entangled pure initial state
built as a superposition, with random phases, of the pure elements that
compose the mixture. This ensures self-averaging of any observable,
drastically reducing the calculation time. The procedure is tested for two
representative systems: a spin star (cluster with random long range
interactions) and a spin ladder.

\end{abstract}
\maketitle

One of the goals of quantum computers is the simulation of quantum systems.
While quantum processing is ideally cast in terms of pure states, experimental
realizations are a major challenge \cite{QCRoadmap04} met only for very small
systems \cite{Koppens06}. The alternative use of statistical mixtures of pure
states, as the spin ensembles in standard NMR experiments
\cite{ZME92,Pastawski95,Madi97,Chuang04}, led to the development of the
ensemble quantum computation (EQC) \cite{EQC} that allowed useful algorithm
optimizations \cite{Long04,Suter05}. Recently, EQC was experimentally
implemented in a $12$-qubits system \cite{Negrevergne06} and much larger
quantum registers \cite{Suter04} were prepared to assess its stability against
decoherence. Hence, an efficient evaluation of ensemble evolutions is needed.
Analytical solutions of ensemble dynamics, either from the integration of the
Liouville von-Neumann equation \cite{Abragam} or the alternative Keldysh
formalism \cite{Keldysh64,Danielewicz84} are limited to special cases (e.g.
\cite{FeldLacelle02,CPL05}). Thus, one has to resort to numerical solutions.
One can describe an ensemble of $M$ spins by using the $2^{M}\times2^{M}%
$\textbf{ }density matrix, but this quickly reaches a storage limit as $M$
increases. Thus, a typical exact diagonalization method in a desktop computer
slightly exceeds a dozen of qubits. Alternatively, the use of wave functions
combined with the Trotter-Suzuki decomposition
\cite{Raedt04,Dobrovitski07} overcomes this limitation because it uses vectors
of size $2^{M}$. However, the average over \emph{individual} evolutions of a
large number of\textbf{ }components of the ensemble takes a long time. Here,
this limitation\ is overcome by profiting of quantum parallelism
\cite{Loss02parallelism} to evaluate the ensemble dynamics of any observable
evolved from a \textquotedblleft local\textquotedblright\ initial condition.
The idea is that when evaluated on \emph{single} pure states that are a
superposition of all the elements of the ensemble, these observables become
self-averaging. This reinforces the suggestions that one can avoid ensemble or
thermal averages by using a single pure state \cite{Popescu06,Rigol08}. Ref.
\cite{Popescu06} considers a subsystem with the reduced density matrix derived
from a pure state, where the subsystem is entangled with an environment which
has a much bigger size. The resulting reduced density matrix describes the
microcanonical ensemble without resorting to the equiprobability postulate of
statistical mechanics. Following a similar inspiration, we focus on the
non-equilibrium dynamics of any given observable. The key is that the initial
non-equilibrium state has a \textquotedblleft local\textquotedblright%
\ character, i.e. starting from an equilibrium ensemble state, a perturbing
excitation acts on a small portion of the system. A number of useful
techniques to simulate dynamics of a small system in the presence of a
mixed-bath that involve averages over random states
\cite{Gelman03,Dobrovitski07} could be interpreted as particular cases.
Specifically, an analytical justification of the numerically observed
self-averaging property in some particular systems \cite{Gelman03} follows
from our results.

While our method is general and can be applied to any mixed many-body system,
we focus on the time evolution of the observable of greatest interest in NMR
experiments, the local polarization \cite{ZME92,Pastawski95,Madi97}. We
consider two spin configurations that are representative of physical
situations of contrasting topology. One is a spin ladder which is a variant of
the linear chains which are exactly solvable \cite{Stolze97,CPL05} and are
known to have strong mesoscopic echoes \cite{Pastawski95,Madi97}. The other is
a spin star, a cluster with random long range interactions which is a fair
representation of many molecular crystals \cite{JCP98}. Here, mesoscopic
interferences becomes less intense \cite{Gruver97}.

\begin{figure}
[tbh]
\begin{center}
\includegraphics[
height=2.5927in,
width=3.4143in
]%
{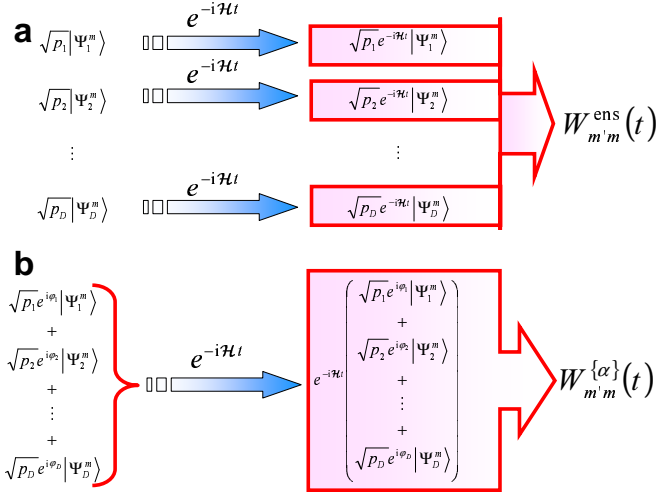}%
\caption{(Color online) Schemes of the quantum evolution of an ensemble (panel
(a)) and a pure-state (panel (b)). Each $\left\vert \Psi_{i}^{m}\right\rangle
=\left\vert \psi_{m}\right\rangle \otimes\left\vert \Psi_{i}\right\rangle $
contains a complete base, $\left\vert \Psi_{i}\right\rangle $, of the $M-m$
spins.}%
\label{Fig_ensemble_vs_entaglement_scheme}%
\end{center}
\end{figure}

\emph{Ensemble vs. pure entangled state.--- }We take the ensemble of all the
many-spin states $\left\vert \Psi_{i}^{m}\right\rangle =\left\vert \psi
_{m}\right\rangle \otimes\left\vert \Psi_{i}\right\rangle $ where $m$ spins
are in the state $\left\vert \psi_{m}\right\rangle $ and the $\left\vert
\Psi_{i}\right\rangle $ are a base for the remaining $M-m$ spins. These states
have a statistical weight $p_{i}$. The probability to find $m^{\prime}$ spins
in the state $\left\vert \psi_{m^{\prime}}\right\rangle $ at time $t$ when the
$m$ spins were in state $\left\vert \psi_{m}\right\rangle $ at $t=0$ is%
\begin{equation}
W_{m^{\prime}m}^{\mathrm{ens}}(t)=\sum_{f=1}^{2^{M-m^{\prime}}}\sum
_{i=1}^{2^{M-m}}p_{i}\left\vert \left\langle \Psi_{f}^{m^{\prime}}\right\vert
e^{-\mathrm{i}\widehat{\mathcal{H}}t/\hbar}\left\vert \Psi_{i}^{m}%
\right\rangle \right\vert ^{2}. \label{Pens}%
\end{equation}
Here, $\left\vert \left\langle \Psi_{f}^{m^{\prime}}\right\vert e^{-\mathrm{i}%
\widehat{\mathcal{H}}t/\hbar}\left\vert \Psi_{i}^{m}\right\rangle \right\vert
^{2}$ is the probability of finding $m^{\prime}$ spins in the state
$\left\vert \psi_{m^{\prime}}\right\rangle $ within the many-spin state
$\left\vert \Psi_{f}^{m^{\prime}}\right\rangle $ at time $t$ provided that, at
$t=0,$ the $m$ spins were in the state $\left\vert \psi_{m}\right\rangle $
within the state $\left\vert \Psi_{i}^{m}\right\rangle $. The sum runs over
all possible initial and final states. An example of this is the local
correlation function where $m=m^{\prime}=1,$ $\left\vert \psi_{m}\right\rangle
=\left\vert \uparrow\right\rangle _{n}$ is the state of the $n$-th spin and
$\left\vert \psi_{m^{\prime}}\right\rangle =\left\vert \uparrow\right\rangle
_{n^{\prime}}$ is the state of spin $n^{\prime}$. The polarization of spin
$n^{\prime}$ at time $t$ provided that the spin $n$ was up at time $t=0$ is
given by $P_{n^{\prime}n}^{\mathrm{ens}}\left(  t\right)  =2[W_{11}%
^{\mathrm{ens}}(t)-1/2]$ \cite{Pastawski95}. The expression (\ref{Pens})
involves $D=2^{M-m}$ different dynamics for each of the initial states, see
Fig. \ref{Fig_ensemble_vs_entaglement_scheme}(a). \textbf{ }%
This number is directly related to the dimension of the Hilbert space. Thus,
the number of computed evolutions increases exponentially with $M$. Our goal
is to extract the same information in a shorter time. The parallelism implicit
in quantum superpositions \cite{Loss02parallelism} suggests that the desired
correlation functions are contained in the dynamics of a \emph{single}
\emph{pure-state}, \textbf{ }see Fig. \ref{Fig_ensemble_vs_entaglement_scheme}%
(b). The pure-state is built as an arbitrary linear superposition of
components of the ensemble, i.e. $\left\vert \Psi_{\mathrm{pure}}^{\{\alpha
\}}\right\rangle =\sum_{i=1}^{D}\alpha_{i}\left\vert \Psi_{i}^{m}\right\rangle
$ where $\alpha_{i}=\sqrt{p_{i}}e^{\mathrm{i}\varphi_{i}}$ with\textbf{
}random $\varphi_{i}$. Thus, the correlation function is given by%
\begin{equation}
W_{m^{\prime}m}^{\{\alpha\}}(t)=%
{\textstyle\sum\limits_{f=1}^{D^{\prime}}}
\left\vert \left\langle \Psi_{f}^{m^{\prime}}\right\vert e^{-\mathrm{i}%
\widehat{\mathcal{H}}t/\hbar}\sum\limits_{i=1}^{D}\alpha_{i}\left\vert
\Psi_{i}^{m}\right\rangle \right\vert ^{2}. \label{Palpha}%
\end{equation}
Here $\{\alpha\}$ denotes the set of all the $\alpha_{i}$ involved in the
initial pure-state and $D^{\prime}=2^{M-m^{\prime}}$. Note that the
substantial difference between Eq. (\ref{Pens}) and Eq. (\ref{Palpha}) is that
the sum on $i$ in the former is outside the square modulus while in the latter
is inside. Rewriting Eq. (\ref{Palpha}) as%
\begin{gather}
W_{m^{\prime}m}^{\{\alpha\}}(t)=%
{\textstyle\sum\limits_{f=1}^{D^{\prime}}}
{\textstyle\sum\limits_{i=1}^{D}}
p_{i}\left\vert \left\langle \Psi_{f}^{m^{\prime}}\right\vert e^{-\mathrm{i}%
\widehat{\mathcal{H}}t/\hbar}\left\vert \Psi_{i}^{m}\right\rangle \right\vert
^{2}\label{Palpha_Cross}\\
+%
{\textstyle\sum\limits_{f=1}^{D^{\prime}}}
{\textstyle\sum\limits_{i\neq i^{\prime}=1}^{D}}
\alpha_{i}^{{}}\alpha_{i^{\prime}}^{\ast}\left\langle \Psi_{i^{\prime}}%
^{m}\right\vert e^{\mathrm{i}\widehat{\mathcal{H}}t/\hbar}\left\vert \Psi
_{f}^{m^{\prime}}\right\rangle \left\langle \Psi_{f}^{m^{\prime}}\right\vert
e^{-\mathrm{i}\widehat{\mathcal{H}}t/\hbar}\left\vert \Psi_{i}^{m}%
\right\rangle ,\nonumber
\end{gather}
the second term contains the initial correlations between the components of
the initial state. These \emph{cross terms }make the difference between
$W_{m^{\prime}m}^{\mathrm{ens}}(t)$ and $W_{m^{\prime}m}^{\{\alpha\}}(t).$ By
performing an extra average over $N_{\alpha}$ realizations of the possible
initial states, we obtain%
\begin{equation}
\left\langle W_{m^{\prime}m}^{\{\alpha\}}(t)\right\rangle _{N_{\alpha}}%
=\frac{1}{N_{\alpha}}%
{\textstyle\sum\limits_{\{\alpha\}}^{N_{\alpha}}}
W_{m^{\prime}m}^{\{\alpha\}}(t). \label{Palpha_aver}%
\end{equation}
Under this average, each element in the cross term goes to zero, hence
$W_{m^{\prime}m}^{\mathrm{ens}}(t)=\lim_{N_{\alpha}\rightarrow\infty
}\left\langle W_{m^{\prime}m}^{\{\alpha\}}(t)\right\rangle _{N_{\alpha}}$.
Thus, by increasing the number $N_{\alpha}$ of evolutions, expression
(\ref{Palpha_aver}) converges to $W_{m^{\prime}m}^{\mathrm{ens}}(t)$ as
described by the central limit theorem. \ The variance of $W_{m^{\prime}%
m}^{\{\alpha\}}(t)$ is $\mathrm{Var}=\lim_{N_{\alpha}\rightarrow\infty
}\left\langle W_{m^{\prime}m}^{\{\alpha\}}(t)^{2}\right\rangle _{N_{\alpha}%
}-\left\langle W_{m^{\prime}m}^{\{\alpha\}}(t)\right\rangle _{N_{\alpha}}%
^{2}.$ At this stage, the four phases $\varphi_{i}$ summing up in each
exponent are correlated. This enables terms where the exponent cancels out.
These survive the average and contribute to\textbf{ }$\mathrm{Var}=~%
{\textstyle\sum\nolimits_{i\neq i^{\prime}=1}^{D}}
p_{i^{{}}}p_{i^{\prime}}\left\vert \left\langle \Psi_{i^{\prime}}%
^{m}\right\vert e^{\mathrm{i}\widehat{\mathcal{H}}t/\hbar}\hat{P}%
_{f}e^{-\mathrm{i}\widehat{\mathcal{H}}t/\hbar}\left\vert \Psi_{i}%
^{m}\right\rangle \right\vert ^{2}$ where $\hat{P}_{f}=%
{\textstyle\sum\nolimits_{f}^{D^{\prime}}}
\left\vert \Psi_{f}^{m^{\prime}}\right\rangle \left\langle \Psi_{f}%
^{m^{\prime}}\right\vert $. In the last expression $\left\vert \left\langle
\Psi_{i^{\prime}}^{m}\right\vert e^{\mathrm{i}\widehat{\mathcal{H}}t/\hbar
}\hat{P}_{f}e^{-\mathrm{i}\widehat{\mathcal{H}}t/\hbar}\left\vert \Psi_{i}%
^{m}\right\rangle \right\vert ^{2}=w_{i^{\prime}i}$ is the probability to find
the state $\left\vert \Psi_{i^{\prime}}^{m}\right\rangle $ after the
subunitary evolution $e^{\mathrm{i}\widehat{\mathcal{H}}t/\hbar}\hat{P}%
_{f}e^{-\mathrm{i}\widehat{\mathcal{H}}t/\hbar}$ provided that the initial
state was $\left\vert \Psi_{i}^{m}\right\rangle .$ The projector $\hat{P}_{f}$
involves only part of the Hilbert space states. Hence, $%
{\textstyle\sum\nolimits_{i^{\prime}}^{D}}
w_{i^{\prime}i}\leq1\ $implies $\mathrm{Var}=$ $%
{\textstyle\sum\nolimits_{\genfrac{}{}{0pt}{}{i,i^{\prime}=1}{i\neq i^{\prime
}}}^{D}}
p_{i^{{}}}p_{i^{\prime}}~w_{i^{\prime}i}\leq$ $%
{\textstyle\sum\nolimits_{i}^{D}}
p_{i}^{{}}\max\left\{  p_{i^{\prime}}\right\}  =$ $\max\left\{  p_{i^{\prime}%
}\right\}  .$ Thus, by using the Chebyshev's Inequality the probability that
$\left\vert \left\langle W_{m^{\prime}m}^{\{\alpha\}}(t)\right\rangle
_{N_{\alpha}}-W_{m^{\prime}m}^{\mathrm{ens}}(t)\right\vert \geq\varepsilon$
(i.e. the error exceeds a desired precision $\varepsilon$) is lower than
$\mathrm{Var}/\left(  N_{\alpha}\varepsilon^{2}\right)  \leq$ $\max\left\{
p_{i^{\prime}}\right\}  /\left(  N_{\alpha}\varepsilon^{2}\right)  .$ For an
homogeneous distribution $p_{i}=1/2^{M-m}$ hence $\max\left\{  p_{i^{\prime}%
}\right\}  /\left(  N_{\alpha}\varepsilon^{2}\right)  =$ $2/\left(
2^{M-m}N_{\alpha}\varepsilon^{2}\right)  .$ The locality of the initial
condition ensures that $M\gg m,$ thus, as $2^{M-m}\ $increases, one gets
$W_{m^{\prime}m}^{\{\alpha\}}(t)\approx W_{m^{\prime}m}^{\mathrm{ens}}(t)$ in
a single realization, i.e. the \emph{cross terms} in Eq. (\ref{Palpha_Cross})
self-average to zero even for $N_{\alpha}=1$.

\emph{Spin systems with different coupling networks.--- }To illustrate the use
of Eq. (\ref{Palpha_aver}) we consider typical situations of high-field
solid-state NMR. Here, the Hamiltonian is simplified by using a frame that
eliminates the Zeeman contribution \cite{Abragam}. We are left with the
spin-spin interaction, $\widehat{\mathcal{H}}=%
{\textstyle\sum\nolimits_{i<j}^{M}}
\left[  a_{ij}\hat{I}_{i}^{z}\hat{I}_{j}^{z}+\tfrac{1}{2}b_{ij}\left(  \hat
{I}_{i}^{+}\hat{I}_{j}^{-}+\hat{I}_{i}^{-}\hat{I}_{j}^{+}\right)  \right]
,$where $b_{ij}/a_{ij}=0$ represents an Ising-like coupling, $a_{ij}/b_{ij}=0$
an $XY$ Hamiltonian, $a_{ij}/b_{ij}=1$ the isotropic one, and $a_{ij}%
/b_{ij}=-2$ a dipolar (secular) Hamiltonian truncated with respect to a Zeeman
field along the $z$ axis.\ The ensemble relevant for NMR experiments is in the
infinite temperature limit \cite{Abragam,Chuang04}, i.e., $p_{i}=1/2^{M-1}%
\ $where $\left\vert \Psi_{i}\right\rangle \ $are simple tensor product states
in the Zeeman basis. The initial conditions are states with a local excitation
at site $n$ over a background level which is determined by the zero
magnetization of the other $M-1$ spins \cite{ZME92,Pastawski95,Madi97}.

\begin{figure}
[tbh]
\begin{center}
\includegraphics[
trim=0.000000in 0.000000in 0.003911in 0.000000in,
height=1.5748in,
width=2.0141in
]%
{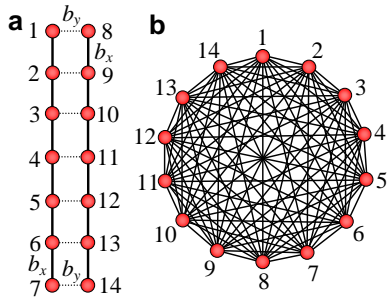}%
\caption{(Color online) Panel (a) shows the coupling network of a spin ladder.
Panel (b) contains the coupling network of a spin star in which all the spins
interact with each other.}%
\label{Fig_ladder_star_diagram}%
\end{center}
\end{figure}

\begin{figure}
[tbh]
\begin{center}
\includegraphics[
height=3.3555in,
width=2.5944in
]%
{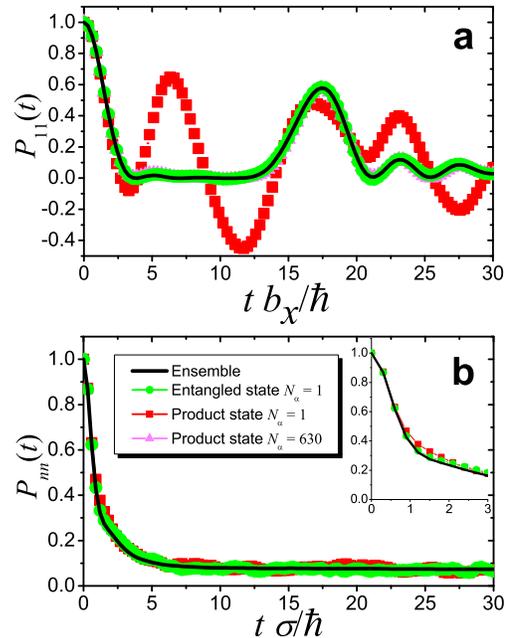}%
\caption{(Color online) Local spin dynamics in a $14$-spin system. The
ensemble dynamics (solid line) is compared with that of entangled and product
pure states. The square (red) and circle (green) scatter points correspond to
$\left\langle P_{11}^{\mathrm{prod}}(t)\right\rangle _{N_{\alpha}}$ and
$\left\langle P_{11}^{\mathrm{ent}}(t)\right\rangle _{N_{\alpha}}$ for
$N_{\alpha}=1$ respectively. (a) The extreme of a spin ladder with
$b_{y}/b_{x}=1/10.$ The triangle scatter points (light magenta) correspond to
$\left\langle P_{11}^{\mathrm{prod}}(t)\right\rangle _{N_{\alpha}}$ with
$N_{\alpha}=630,$ the lower value yielding the ensemble dynamics. A strong
mesoscopic echo is evident. (b) A site in a spin star with random dipolar
interactions. No mesoscopic echo is evident. }%
\label{Fig_Ent_vs_Ens}%
\end{center}
\end{figure}

We calculate the local polarization of site $n^{\prime}$ at time $t$ provided
that it was polarized ($n=n^{\prime}$)\ at time $t=0$ in two different spin
systems which have well differentiated kinds of dynamics:

a) A \emph{ladder} of spins interacting through an XY Hamiltonian, as shown in
Fig. \ref{Fig_ladder_star_diagram}(a). There, $a_{ij}=0,$ $b_{i,i+1}%
=b_{i+M/2,i+M/2+1}=b_{x}$ and $b_{i,i+M/2}=b_{y}$. Here, the exact dynamics
presents long lived recurrences (mesoscopic echoes) shown by the black line in
Fig. \ref{Fig_Ent_vs_Ens}(a), due to the high symmetry in the coupling
topology \cite{Pastawski95,Madi97}. The method also reproduces the exact
solutions in isolated spin chains with both, pure XY or XY plus Ising
interactions \cite{Stolze97}. Moreover, the results confirm that inclusion of
Ising terms or interchain couplings leads to decoherence degrading the
mesoscopic echoes \cite{Madi97}.

b)\emph{ }A \emph{star }system, see Fig. \ref{Fig_ladder_star_diagram}(b), in
which all the spins interact with each other through a dipolar coupling
$a_{ij}/b_{ij}=-2$. The coupling intensities are given by a Gaussian random
distribution with zero mean and variance $\sigma^{2}$. In this case, the local
polarization decays with a rate proportional to the square root of the
\emph{local }second moment $\sigma_{0}^{2}=\frac{9}{4}\left(  M-1\right)
\sigma^{2}$ \cite{Abragam} of the Hamiltonian and recurrences are negligible
\cite{Gruver97}. The black line (exact solution) of Fig. \ref{Fig_Ent_vs_Ens}%
(b) shows the local polarization of this system.

\emph{Testing the quantum parallelism.--- }In order to compare Eq.
(\ref{Palpha_aver}) to the ensemble average of Eq. (\ref{Pens}), as well as
its dependence on the choice of the phases $\varphi_{i}$, we calculate the
evolution for two types of initial states. Firstly, a pure entangled state is
constructed by choosing $\varphi_{i}$ randomly. Thus, assuming $n=1,$
$\left\vert \Psi_{\mathrm{pure}}^{\{\alpha\}}\right\rangle $ becomes
$\left\vert \Psi_{\mathrm{ent}}^{\{\alpha\}}\right\rangle =$ $%
{\textstyle\sum_{i=1}^{2^{M-1}}}
\tfrac{1}{\sqrt{2^{M-1}}}e^{-\mathrm{i}\varphi_{i}}\left\vert \uparrow
\right\rangle _{1}\otimes\left\vert \Psi_{i}\right\rangle .$ The correlation
function, Eq. (\ref{Palpha_aver}), calculated with this state is $\left\langle
W_{11}^{\mathrm{ent}}(t)\right\rangle _{N_{\alpha}}$ giving the polarization
$\left\langle P_{11}^{\mathrm{ent}}(t)\right\rangle _{N_{\alpha}}=$ $2\left(
\left\langle W_{11}^{\mathrm{ent}}(t)\right\rangle _{N_{\alpha}}-1/2\right)
$. The second case is a product (not entangled) state. It is built with the
$n$-th spin \emph{up} and all the others in a linear combination of spins
\emph{up} and \emph{down,} with equal probability and arbitrary phase.
Assuming $n=1$, we have $\left\vert \Psi_{\mathrm{prod}}^{\{\alpha
\}}\right\rangle =$ $\left\vert \uparrow\right\rangle _{1}%
{\textstyle\bigotimes}
\prod\limits_{l=2}^{M}\left\vert \rightarrow\right\rangle _{l},$ where
$\left\vert \rightarrow\right\rangle _{l}=$ $\tfrac{1}{\sqrt{2}}\left(
\left\vert \downarrow\right\rangle _{l}+\left\vert \uparrow\right\rangle
_{l}e^{-\mathrm{i}\phi_{l}}\right)  $ with $\phi_{l}$\ random variables. Note
that this state can be rewritten in the form of $\left\vert \Psi
_{\mathrm{pure}}^{\{\alpha\}}\right\rangle $ where the resulting phases
$\varphi_{i}$ are correlated. Here, the correlation function
(\ref{Palpha_aver}) is $\left\langle W_{11}^{\mathrm{prod}}(t)\right\rangle
_{N_{\alpha}}$ and the polarization is $\left\langle P_{11}^{\mathrm{prod}%
}(t)\right\rangle _{N_{\alpha}}=$ $2\left\langle W_{11}^{\mathrm{prod}%
}(t)\right\rangle _{N_{\alpha}}-1.$

The local polarization, $P_{11}^{\mathrm{ens}}(t)=2\left(  W_{11}%
^{\mathrm{ens}}(t)-1/2\right)  ,$ obtained with Eq. (\ref{Pens}), for the
$14$-spin ladder system is shown in Fig. \ref{Fig_Ent_vs_Ens}(a) with\ a black
line.
Square (red) and circle (green) scatter lines correspond to the temporal
evolution of $\left\langle P_{11}^{\mathrm{prod}}(t)\right\rangle _{N_{\alpha
}}$ and $\left\langle P_{11}^{\mathrm{ent}}(t)\right\rangle _{N_{\alpha}}%
\ $respectively, with $N_{\alpha}=1$. The agreement between $P_{11}%
^{\mathrm{ens}}(t)$ and $\left\langle P_{11}^{\mathrm{ent}}(t)\right\rangle
_{1}$ is excellent, while $\left\langle P_{11}^{\mathrm{prod}}(t)\right\rangle
_{1}$ has a dynamics quite different from that of the ensemble. The difference
between the dynamics of the two initial pure states is due to the different
number of independent random phases of each state. In the random entangled
state, there are $2^{M-1}$ independent phases that make the cancellation of
the second term in the rhs of Eq. (\ref{Palpha_Cross}) possible. However, the
number of independent phases for the product state is $M-1$. This implies that
there are multiple correlations between the phases in the \emph{cross terms
}inhibiting their self cancellation.
The triangle (light magenta) scatter line in Fig. \ref{Fig_Ent_vs_Ens}(a)
shows the dynamics of $\left\langle P_{11}^{\mathrm{prod}}(t)\right\rangle
_{N_{\alpha}}.$ It becomes indistinguishable from the exact dynamics provided
that $N_{\alpha}\gtrsim630$. The relation between $N_{\alpha}^{\mathrm{prod}%
}=630$ and $N_{\alpha}^{\mathrm{ent}}=1$ is determined by the number of
independent phases associated with the dimension of the sampled portion of the
Hilbert space \cite{Popescu06}, i.e. $8192=N_{\alpha}^{\mathrm{ent}}%
2^{M-1}\simeq N_{\alpha}^{\mathrm{prod}}(M-1)=8190.$

Fig. \ref{Fig_Ent_vs_Ens}(b) shows the local polarization for the spin star
system. The complexity of this system washes out any possible recurrence for
long times leading to a form of spin \textquotedblleft
diffusion\textquotedblright. For $N_{\alpha}=1$ both $\left\langle
P_{11}^{\mathrm{prod}}(t)\right\rangle _{1}$ and $\left\langle P_{11}%
^{\mathrm{ent}}(t)\right\rangle _{1}$ are almost indistinguishable from the
ensemble dynamics. This contrasts with the spin ladder where one would need
$\left\langle P_{11}^{\mathrm{prod}}(t)\right\rangle _{630}$ to get a fair
description. Notably, $\left\langle P_{11}^{\mathrm{ent}}(t)\right\rangle
_{1}$ is an excelent approximant of the ensemble for both cases. This is
because, in the star system the cross terms $\left\langle \Psi_{f}%
^{1}\right\vert e^{-\mathrm{i}\widehat{\mathcal{H}}t/\hbar}\left\vert \Psi
_{i}^{1}\right\rangle \left\langle \Psi_{i^{\prime}}^{1}\right\vert
e^{\mathrm{i}\widehat{\mathcal{H}}t/\hbar}\left\vert \Psi_{f}^{1}\right\rangle
$ of Eq. (\ref{Palpha_Cross}) decay to a value of the order of $1/2^{M}$
within a time scale determined by the Hamiltonian second moment $\sigma
_{\mathcal{H}}^{2}=\frac{M}{2}\sigma_{0}^{2}.$ Thus, even the few $M$
independent phases are enough to cancel the cross terms. In contrast, in the
ladder system, the terms $\left\langle \Psi_{f}^{1}\right\vert e^{-\mathrm{i}%
\widehat{\mathcal{H}}t/\hbar}\left\vert \Psi_{i}^{1}\right\rangle \left\langle
\Psi_{i^{\prime}}^{1}\right\vert e^{\mathrm{i}\widehat{\mathcal{H}}t/\hbar
}\left\vert \Psi_{f}^{1}\right\rangle $ present strong correlations and thus,
the role of the phases becomes more relevant.

In summary, we developed a method to overcome the limitations of the numerical
calculations of an ensemble spin dynamics for large number of spins. Instead
of evolving every one of the $2^{M-m}$ initial states, when $2^{M}\gg2^{m},$
we evolve a single random entangled state. The procedure exploits the quantum
parallelism implicit in quantum superpositions \cite{Loss02parallelism} to
reproduce the ensemble dynamics of any observable. This result supports a
novel view of the foundation of equilibrium statistical mechanics
\cite{Popescu06,Rigol08}. Moreover, even the non-equilibrium statistical
theory of the density matrix describing an ensemble in the thermodynamic limit
could now be based on single states. Here, we observe that even for systems as
small as $14$ spins, the equivalence between a randomly correlated pure-state
and an ensemble state holds. This is a consequence of the exponential increase
of the dimension of the Hilbert space with the system size. The power of the
method is enhanced when combined with the Trotter-Suzuki decomposition. We
showed that the contribution of the extra correlations of the initial
pure-state to the dynamics becomes negligible by increasing $2^{M-m}$, the
ratio between the size of the system Hilbert space and that of the subsystem
where the non-equilibrium initial condition is supported. The method developed
here allows for very efficient dynamical calculations of common experimental
situations where large ensembles are involved. Conversely, it prescribes
possible pure input states for a quantum simulator to yield ensemble evolutions.

\begin{acknowledgments}
We acknowledge support from Fundaci\'{o}n Antorchas, CONICET, FoNCyT, and
SeCyT-UNC. G.A.A. is a postdoctoral fellow of CONICET. E.P.D. thanks the
Alexander von Humboldt Foundation for a Research Scientist Fellowship. P.R.L.
and H.M.P. are members of the Research Career of CONICET. This work has
benefited from discussions with G.A. Raggio, J.P. Paz, F.M. Cucchietti, G.Usaj
as well as very fruitful comments from F.M. Pastawski.
\end{acknowledgments}

\bibliographystyle{apsrev}
\bibliography{biblio_resumida_articles}

\end{document}